\documentclass[10pt,a4paper]{article}
\usepackage{amsmath}
\usepackage{amssymb}
\usepackage{amsfonts}
\usepackage{amstext}
\usepackage{amsbsy}
\usepackage[mathscr]{eucal}
\usepackage{graphicx}
\usepackage{color}
\usepackage[all]{xy}
\newcommand{\beq}{\begin{equation}}
\newcommand{\eeq}{\end{equation}}
\newcommand{\beqa}{\begin{eqnarray}}
\newcommand{\eeqa}{\end{eqnarray}}
\newcommand{\ket}[1]{| #1 \rangle}

\newtheorem{thm}{Theorem}[subsection]
 
 \newtheorem{lem}[thm]{Lemma}

 \numberwithin{equation}{subsection}

\makeindex
\title{\Large\textbf{Topological quantum gate entangler for a multi-qubit state}}

\author{\textit{ Hoshang Heydari}\\
        \small\textit{Institute of Quantum
Science, Nihon University,}\\
\small\textit{1-8 Kanda-Surugadai, Chiyoda-ku, Tokyo 101-8308,
Japan}
\\\small\textit{Email: hoshang@edu.cst.nihon-u.ac.jp}}

\date{}
\begin{document}

\maketitle \thispagestyle{empty}

\begin{abstract}
We establish a relation between topological and quantum
entanglement for a multi-qubit state by considering the unitary
representations of the Artin braid group. We construct topological
operators that can entangle multi-qubit state. In particular we
construct operators that create quantum entanglement for
multi-qubit states based on the Segre ideal of complex
multi-projective space. We also in detail discuss and construct
these operators for two-qubit and three-qubit states.
\end{abstract}

\section{Introduction}
Multipartite entangled states are the building block of a universal
quantum computer. For example an one-way quantum computer as a scheme
for universal quantum computation are based on entangled cluster
states. Recently, L.  Kauffman and S.  Lomonaco Jr. have shown
that topological entanglement and quantum entanglement are closely
related \cite{kauf}. They introduced a topological operator called
braiding operator that can entangle quantum state. These operator
are solution of Yang-Baxter equation. The braiding operator are
also unitary transformation which make them very suitable for
application in the field of quantum computing.  We have also
recently establish a relation between multipartite states and the
Segre variety and the Segre ideal \cite{Hosh5,Hosh6}. For example,
we have shown that the Segre ideal represent completely separable
states of multipartite states.   In this paper, we will construct
braiding operators for multi-qubit states based on construction of
the Segre ideal. In particular, in section \ref{Com} we will give
a short introduction to complex projective variety and complex
multi-projective Segre variety and ideal. In section \ref{Top} we
will review the basic construction of topological entanglement
operators. We also discuss braiding operator for two-qubit state.
Finally in section \ref{mul} we will construct such topological
unitary operators for multi-qubit states. We will in detail
discuss the construction of this operator for a three-qubit state.
Now, denote a general, composite quantum system with $m$
subsystems as
$\mathcal{Q}=\mathcal{Q}^{p}_{m}(N_{1},N_{2},\ldots,N_{m})
=\mathcal{Q}_{1}\mathcal{Q}_{2}\cdots\mathcal{Q}_{m}$, with the
pure state $
\ket{\Psi}=\sum^{N_{1},N_{2},\ldots,N_{m}}_{k_{1},k_{2},\ldots,k_{m}=1}\alpha_{k_{1}k_{2}\ldots
k_{m}} \ket{k_{1}k_{2}\ldots k_{m}} $ and corresponding Hilbert
space $
\mathcal{H}_{\mathcal{Q}}=\mathcal{H}_{\mathcal{Q}_{1}}\otimes
\mathcal{H}_{\mathcal{Q}_{2}}\otimes\cdots\otimes\mathcal{H}_{\mathcal{Q}_{m}}
$, where the dimension of the $j$th Hilbert space is
$N_{j}=\dim(\mathcal{H}_{\mathcal{Q}_{j}})$. We are going to use
this notation throughout this paper. In particular, we denote a
pure two-qubit state by $\mathcal{Q}^{p}_{2}(2,2)$. Next, let
$\rho_{\mathcal{Q}}$ denote a density operator acting on
$\mathcal{H}_{\mathcal{Q}}$. The density operator
$\rho_{\mathcal{Q}}$ is said to be fully separable, which we will
denote by $\rho^{sep}_{\mathcal{Q}}$, with respect to the Hilbert
space decomposition, if it can  be written as $
\rho^{sep}_{\mathcal{Q}}=\sum^\mathrm{N}_{k=1}p_k
\bigotimes^m_{j=1}\rho^k_{\mathcal{Q}_{j}},~\sum^\mathrm{N}_{k=1}p_{k}=1
$
 for some positive integer $\mathrm{N}$, where $p_{k}$ are positive real
numbers and $\rho^k_{\mathcal{Q}_{j}}$ denotes a density operator on
Hilbert space $\mathcal{H}_{\mathcal{Q}_{j}}$. If
$\rho^{p}_{\mathcal{Q}}$ represents a pure state, then the quantum
system is fully separable if $\rho^{p}_{\mathcal{Q}}$ can be written
as
$\rho^{sep}_{\mathcal{Q}}=\bigotimes^m_{j=1}\rho_{\mathcal{Q}_{j}}$,
where $\rho_{\mathcal{Q}_{j}}$ is the density operator on
$\mathcal{H}_{\mathcal{Q}_{j}}$. If a state is not separable, then
it is said to be an entangled state.
\section{Complex projective variety and Segre ideal for multi-qubit
state}\label{Com} In this section,
 we will define complex projective space, ideal, and variety.
 Moreover, we will review the construction of the
Segre ideal for multi-qubit state. Here are some general
references on complex projective  geometry  \cite{Griff78,Mum76}.
A complex projective space $\mathbb{P}_{\mathbb{C}}^{n}$ is
defined to be the set of lines through the origin in
$\mathbb{C}^{n+1}$, that is, $
\mathbb{P}_{\mathbb{C}}^{n}=\frac{\mathbb{C}^{n+1}-{0}}{
(x_{1},\ldots,x_{n+1})\sim(y_{1},\ldots,y_{n+1})},~\lambda\in
\mathbb{C}-0,~y_{i}=\lambda x_{i} ~\forall ~0\leq i\leq n+1. $
Let $C[z]=C[z_{1},z_{2}, \ldots,z_{n}]$ denotes the polynomial
algebra in $n$  variables with complex coefficients. Then, given a
set of homogeneous polynomials $\{g_{1},g_{2},\ldots,g_{q}\}$ with
$g_{i}\in C[z]$, we define a complex projective variety as
\begin{eqnarray}
&&\mathcal{V}(g_{1},\ldots,g_{q})=\{O\in\mathbb{P}_{\mathbb{C}}^{n}:
g_{i}(O)=0~\forall~1\leq i\leq q\},
\end{eqnarray}
where $O=[a_{1},a_{2},\ldots,a_{n+1}]$ denotes the equivalent
class of point $\{\alpha_{1},\alpha_{2},\ldots,$
$\alpha_{n+1}\}\in\mathbb{C}^{n+1}$. Let $\mathcal{V}$ be complex
projective variety. Then an ideal of $\mathbf{C}[z_{1},z_{2},
\ldots,z_{n}]$ is defined by
\begin{eqnarray}
&&\mathcal{I}(\mathcal{V})=\{g\in\mathbf{C}[z_{1},z_{2},
\ldots,z_{n}]: g(z)=0~\forall~z\in\mathcal{V}\}.
\end{eqnarray}
 Note also that
 $\mathcal{V}(\mathcal{I}(\mathcal{V}))=\mathcal{V}$.
  We can map the product of  spaces
$\mathbb{P}_{\mathbb{C}}^{N_{1}-1}\times\mathbb{P}_{\mathbb{C}}^{N_{2}-1}
\times\cdots\times\mathbb{P}_{\mathbb{C}}^{N_{m}-1}$ into a
projective space by its Segre embedding as follows. Let
$(\alpha^{i}_{1},\alpha^{i}_{2},\ldots,\alpha^{i}_{N_{i}})$  be
points defined on the complex projective space
$\mathbb{P}_{\mathbb{C}}^{N_{i}-1}$. Then the Segre map
\begin{equation}\xymatrix{\mathbb{P}_{\mathbb{C}}^{N_{1}-1}\times\mathbb{P}_{\mathbb{C}}^{N_{2}-1}
\times\cdots\times\mathbb{P}_{\mathbb{C}}^{N_{m}-1}
\ar[r]_{~~\mathcal{S}_{N_{1},\ldots,N_{m}}}&\mathbb{P}_{\mathbb{C}}^{N_{1}N_{2}\cdots
N_{m}-1}}
\end{equation}
is defined by $
 ((\alpha^{1}_{1},\alpha^{1}_{2},\ldots,\alpha^{1}_{N_{1}}),\ldots,
 (\alpha^{m}_{1},\alpha^{m}_{2},\ldots,\alpha^{m}_{N_{m}}))$ $  \longmapsto
 (\alpha^{1}_{i_{1}}\alpha^{2}_{i_{2}}\cdots\alpha^{m}_{i_{m}})$.
Next, let $\alpha_{i_{1}i_{2}\cdots i_{m}}$,$1\leq i_{j}\leq
N_{j}$ be a homogeneous coordinate-function on
$\mathbb{P}_{\mathbb{C}}^{N_{1}N_{2}\cdots N_{m}-1}$. For a
multi-qubit quantum system the Segre ideal is defined by
\begin{equation}\label{segr}
\mathcal{I}^{m}_{\mathrm{Segre}}=\sum^{m}_{j=1}\mathcal{I}_{\mathcal{Q}_{j}\models\mathcal{Q}_{1}\mathcal{Q}_{2}
\cdots\widehat{\mathcal{Q}}_{j}\cdots\mathcal{Q}_{m}},
\end{equation} where
$\mathcal{I}_{\mathcal{Q}_{j}\models\mathcal{Q}_{1}\mathcal{Q}_{2}
\cdots\widehat{\mathcal{Q}}_{j}\cdots\mathcal{Q}_{m}}$ is the
ideal defining when a subsystem $\mathcal{Q}_{j}$ is separated
from quantum system $\mathcal{Q}_{1}\mathcal{Q}_{2}
\cdots\mathcal{Q}_{m}$ is generated by
\begin{eqnarray}
&&\mathcal{I}_{\mathcal{Q}_{j}\models\mathcal{Q}_{1}\mathcal{Q}_{2}\cdots
\widehat{\mathcal{Q}}_{j}\cdots\mathcal{Q}_{m}}=\left\langle\text{Minors}_{2\times
2}\mathcal{X}^{j}_{2\times 2^{m-1}}\right\rangle,
\end{eqnarray}
where $\mathcal{X}^{j}_{2\times 2^{m-1}}$ is the following
$2\times 2^{m-1}$ matrix
\begin{equation}\left(%
\begin{array}{cccc}
  \alpha_{11\ldots11_{j}1\ldots1} & \alpha_{11\ldots11_{j}1\ldots2} & \ldots & \alpha_{22\ldots21_{j}2\ldots2}\\
  \alpha_{11\ldots12_{j}1\ldots1} & \alpha_{11\ldots12_{j}1\ldots2} & \ldots & \alpha_{22\ldots22_{j}2\ldots2}\\
\end{array}%
\right).
\end{equation}
where $j=1,2,\ldots,m$ and
$\mathcal{Q}_{j}\models\mathcal{Q}_{1}\mathcal{Q}_{2}
\cdots\widehat{\mathcal{Q}}_{j}\cdots\mathcal{Q}_{m}$ means we
delete $\mathcal{Q}_{j}$ from right side and add it to the left
side of $\models$.

\section{Topological entanglement operators}\label{Top}
In this section we will give a short introduction to Artin braid
group and Yang-Baxter equation. We will study relation between
topological and quantum entanglement by investigating the unitary
representation of Artin braid group. Here are some general
references on quantum group and low-dimensional topology
\cite{kassel,chari}. The Artin braid group $\mathrm{B}_{n}$ on $n$
strands is generated by $\{b_{n}:1\leq i\leq n-1\}$ and we have
the following relations in the group $\mathrm{B}_{n}$: i) $
    b_{i}b_{j}=b_{j}b_{i}$ for $|i-j|\geq n $
and
 ii)  $b_{i}b_{i+1}b_{i}=b_{i+1}b_{i}b_{i+1}$ for $ 1\leq i<n$.
 Let $\mathcal{V}$ be a complex vector space. Then, for two
strand braid there is associated an operator
$\mathcal{R}:\mathcal{V}\otimes\mathcal{V}\longrightarrow\mathcal{V}\otimes\mathcal{V}$.
Moreover, let $\mathcal{I}$ be the identity operator on
$\mathcal{V}$. Then, the Yang-Baxter equation is defined by
\begin{equation}\label{YB}
    (\mathcal{R}\otimes \mathcal{I})(\mathcal{I}\otimes \mathcal{R})(\mathcal{R}\otimes
    \mathcal{I})=(\mathcal{I}\otimes \mathcal{R})(\mathcal{R}\otimes \mathcal{I})(\mathcal{I}\otimes
    \mathcal{R}).
\end{equation}
The Yang-Baxter equation represents the fundamental topological
relation in the Artin braid group. The inverse to $\mathcal{R}$ will
be associated with the reverse elementary braid on two strands.
Next, we define a representation $\tau$ of the Artin braid group to
the automorphism of $\mathcal{V}^{\otimes
m}=\mathcal{V}\otimes\mathcal{V}\otimes\cdots\otimes\mathcal{V}$ by
\begin{equation}\label{ABG}
    \tau(b_{i})=\mathcal{I}\otimes\cdots\otimes\mathcal{I}\otimes\mathcal{R}\otimes\mathcal{I}\otimes\cdots\otimes\mathcal{I},
\end{equation}
where $\mathcal{R}$ are in position $i$ and $i+1$. This equation
describe a representation of the braid group if $\mathcal{R}$
satisfies the Yang-Baxter equation and is also invertible.
Moreover, this representation of braid group is unitary if
$\mathcal{R}$ is also unitary operator. Thus $\mathcal{R}$ being
unitary indicated that this operator can performs topological
entanglement and it also can be considers as quantum gate. It has
been show in \cite{kauf} that $\mathcal{R}$ can also perform
quantum entanglement by acting on qubits states. Now, let
$\alpha_{11},\alpha_{12},\alpha_{21}$, and $\alpha_{22}$ be any
scalars on the unit circle in the complex plane. Then, we  can
construct an unitary $\mathcal{R}$ as follow
\begin{equation}\label{R}
\mathcal{R}=\left(
  \begin{array}{cccc}
    \alpha_{11} & 0& 0 & 0 \\
    0 & 0 & \alpha_{12} & 0 \\
    0 & \alpha_{21} & 0 &0 \\
    0 & 0 & 0 & \alpha_{22} \\
  \end{array}
\right)
\end{equation}
which is a solution to the Yang-Baxter equation. To see how it is
related to quantum gates, let $\mathcal{P}$ be the swap gate
$\tau=\mathcal{R}\mathcal{P}$ gate  define by
\begin{equation}\label{R}
\mathcal{P}=\left(
  \begin{array}{cccc}
    1 & 0& 0 & 0 \\
    0 & 0 & 1 & 0 \\
    0 & 1 & 0 &0 \\
    0 & 0 & 0 & 1 \\
  \end{array}
\right),~\tau=\left(
  \begin{array}{cccc}
    \alpha_{11} & 0& 0 & 0 \\
    0 &  \alpha_{12}& & 0 \\
    0 & 0&\alpha_{21}  &0 \\
    0 & 0 & 0 & \alpha_{22} \\
  \end{array}
\right)
\end{equation}
In view of braiding and algebra, $\mathcal{R}$ is a solution to
the braided version of the Yang-Baxter equation and $\tau$ is a
solution to the algebraic Yang-Baxter equation and $\mathcal{P}$
represent a virtual or flat crossing. The action of unitary matrix
$\mathcal{R}$ on a quantum state are: i)
$\mathcal{R}\ket{11}=\alpha_{11}\ket{11}$, ii)
$\mathcal{R}\ket{12}=\alpha_{21}\ket{21}$, iii)
$\mathcal{R}\ket{21}=\alpha_{12}\ket{12}$, iv)
$\mathcal{R}\ket{22}=\alpha_{22}\ket{22}$. A proof that the
operator $\mathcal{R}$ can entangle quantum states is give in
\cite{kauf}. Here, we will also give a proof based on the
construction of the Segre variety.
\begin{lem} If elements of $\mathcal{R}$ satisfies
$\alpha_{11}\alpha_{22}\neq\alpha_{12}\alpha_{21}$, then the state
$\mathcal{R}(\ket{\psi}\otimes\ket{\psi})$, with
$\ket{\psi}=\ket{1}+\ket{2}$ is entangled.
\end{lem}
From the construction of the Segre ideal the separable set of two
qubit state satisfies
$\alpha_{11}\alpha_{22}=\alpha_{12}\alpha_{21}$. Thus a two qubit
state
\begin{equation}\label{TQ}
\mathcal{R}(\ket{\psi}\otimes\ket{\psi})=\alpha_{11}\ket{11}
   +\alpha_{12}\ket{12}+\alpha_{21}\ket{21}+\alpha_{22}\ket{22}.
\end{equation}
 is
entangled if and only if  this inequality does not hold.
 We can
also note that a measure of entanglement for two-qubit state in
give by concurrence
$\mathcal{C}(\ket{\Phi})=2|\alpha_{11}\alpha_{22}-\alpha_{12}\alpha_{21}|$.
In general, let $M=(\mathcal{M}_{kl})$ denote an $n\times n$
matrix with complex elements and let $\mathcal{R}$ be defined by
$\mathcal{R}^{kl}_{rs}=\delta^{k}_{s}\delta^{l}_{r}\mathcal{M}_{kl}$.
Then $\mathcal{R}$ is a unitary solution to the Yang-Baxter
equation.
 In the next section, we will used this construction and proof to
create entangled states for three-qubit states.
\section{Multi-qubit quantum gate entangler}\label{mul}
 In
the previous section we have shown that how we can create
entangled state using topological unitary transformation
$\mathcal{R}$. We have also show a relation between the Segre
ideal and such transformation. In this section, we will use this
information to construct multi-qubit entangled state based on this
ideal. But first, we will construct such topological operator for
three-qubit state. For this state the ideal
$\mathcal{I}^{2,2,2}_{\mathcal{Q}_{1}\models\mathcal{Q}_{2}\mathcal{Q}_{3}}$
representing if a subsystem $\mathcal{Q}_{1}$ that is unentangled
with  $\mathcal{Q}_{2}\mathcal{Q}_{3}$ is generated by
$\mathcal{I}_{\mathcal{Q}_{1}\models\mathcal{Q}_{2}\mathcal{Q}_{3}}=\left\langle\text{Minors}_{2\times
2}\mathcal{X}^{1}_{2\times 4}\right\rangle$, that is
\begin{equation}
\mathcal{I}_{\mathcal{Q}_{1}\models\mathcal{Q}_{2}\mathcal{Q}_{3}}=\left\langle\text{Minors}_{2\times 2}\left(%
\begin{array}{cccc}
  \alpha_{111} & \alpha_{112}&\alpha_{121}&\alpha_{122} \\
 \alpha_{211} & \alpha_{212}&\alpha_{221}&\alpha_{222} \\
\end{array}%
\right)\right\rangle,
\end{equation}
where we have used the notation $\models$ to indicate that
$\mathcal{Q}_{1}$ is separated from
$\mathcal{Q}_{2}\mathcal{Q}_{3}$ but we still could have
entanglement between $\mathcal{Q}_{2}$ and $\mathcal{Q}_{3}$.
In the same way, we can define the ideal
$\mathcal{I}^{2,2,2}_{\mathcal{Q}_{2}\models\mathcal{Q}_{1}\mathcal{Q}_{3}}$
representing if the subsystem $\mathcal{Q}_{2}$ is unentangled
with $\mathcal{Q}_{1}\mathcal{Q}_{3}$ and
$\mathcal{I}_{\mathcal{Q}_{3}\models\mathcal{Q}_{1}\mathcal{Q}_{2}}$
representing if the subsystem $\mathcal{Q}_{3}$ is unentangled
with $\mathcal{Q}_{2}\mathcal{Q}_{3}$. The ideals are generated $
\mathcal{I}_{\mathcal{Q}_{2}\models\mathcal{Q}_{1}\mathcal{Q}_{3}}=\left\langle\text{Minors}_{2\times
2}\mathcal{X}^{2}_{2\times 4}\right\rangle,~ \text{and}~
\mathcal{I}_{\mathcal{Q}_{3}\models\mathcal{Q}_{1}\mathcal{Q}_{2}}=\left\langle\text{Minors}_{2\times
2}\mathcal{X}^{3}_{2\times 4}\right\rangle.$
 Thus, the Segre ideal
for three-qubit state is given by
\begin{eqnarray}
  \mathcal{I}^{3}_{\mathrm{Segre}}&=& \mathcal{I}_{\mathcal{Q}_{1}\models\mathcal{Q}_{2}\mathcal{Q}_{3}}
+\mathcal{I}_{\mathcal{Q}_{2}\models\mathcal{Q}_{1}\mathcal{Q}_{3}}
+\mathcal{I}_{\mathcal{Q}_{3}\models\mathcal{Q}_{1}\mathcal{Q}_{2}}\\\nonumber&=&
\left\langle\mathrm{T}_{1},\mathrm{T}_{2},\ldots,\mathrm{T}_{12}\right\rangle,
\end{eqnarray}
where
$\mathrm{T}_{1}=\alpha_{1,1,1}\alpha_{2,2,1}-\alpha_{1,2,1}\alpha_{2,1,1}$,
$\mathrm{T}_{2}=\alpha_{1,1,2}\alpha_{2,2,2}-\alpha_{1,2,2}\alpha_{2,1,2}$,
$\mathrm{T}_{3}=\alpha_{1,1,1}\alpha_{2,1,2}-\alpha_{1,1,2}\alpha_{2,1,1}$,
$\mathrm{T}_{4}=\alpha_{1,2,1}\alpha_{2,2,2}-\alpha_{1,2,2}\alpha_{2,2,1}$,
$\mathrm{T}_{5}=\alpha_{1,1,1}\alpha_{1,2,2}-\alpha_{1,1,2}\alpha_{1,2,1}$,
$\mathrm{T}_{6}=\alpha_{2,1,1}\alpha_{2,2,2}-\alpha_{2,1,2}\alpha_{2,2,1}$,
$\mathrm{T}_{7}=\alpha_{1,1,1}\alpha_{2,2,2}-\alpha_{1,1,2}\alpha_{2,2,1}$,
$\mathrm{T}_{8}=\alpha_{1,1,1}\alpha_{2,2,2}-\alpha_{1,2,1}\alpha_{2,1,2}$,
$\mathrm{T}_{9}=\alpha_{1,1,1}\alpha_{2,2,2}-\alpha_{1,2,2}\alpha_{2,1,1}$,
$\mathrm{T}_{10}=\alpha_{1,1,2}\alpha_{2,2,1}-\alpha_{1,2,1}\alpha_{2,1,2}$,
$\mathrm{T}_{11}=\alpha_{1,2,1}\alpha_{2,1,2}-\alpha_{1,2,2}\alpha_{2,1,1}$,
and
$\mathrm{T}_{12}=\alpha_{1,2,1}\alpha_{2,1,2}-\alpha_{1,2,2}\alpha_{2,1,1}$.
In our recent paper \cite{Hosh5} we have shown that we can
construct a measure of entanglement for three-qubit states based
on these Segre varieties. We have also construct a measure of
entanglement for general multipartite states based on an extension
of the Segre varieties \cite{Hosh6}. For example, for three-qubit
state a measure of entanglement is given by
\begin{eqnarray}\nonumber
\mathcal{C}(\ket{\Psi})&=&( 2(| \mathrm{T}_{1}|^{2} +
|\mathrm{T}_{2}|^{2}
 +|\mathrm{T}_{3}|^{2}+
 |\mathrm{T}_{4}|^{2}+
|\mathrm{T}_{5}|^{2}+| \mathrm{T}_{6}|^{2})\\\nonumber&&  + |\mathrm{T}_{7}
|^{2}
+ | \mathrm{T}_{8} |^{2}  + |\mathrm{T}_{9}|^{2}
+ | \mathrm{T}_{10}|^{2}+| \mathrm{T}_{11}|^{2}+|
\mathrm{T}_{12}|^{2})^{\frac{1}{2}},
\end{eqnarray}
Now, based on comparison with the two-qubit case we will construct
a unitary transformation $\mathcal{R}$ that create three-qubit
entangled states. Let
\begin{equation}\mathcal{R}=\left(%
\begin{array}{cccccccc}
  \alpha_{111} & 0 & 0 & 0 & 0 & 0 & 0 & 0 \\
  0 & 0 & 0 & 0 & 0 & 0 & \alpha_{221} & 0 \\
  0 & 0 & 0 & 0 & 0 & \alpha_{212}& 0 & 0 \\
  0 & 0 & 0 & 0 & \alpha_{211} & 0 & 0 & 0 \\
   0 & 0 & 0 & \alpha_{122} & 0 & 0 & 0 & 0 \\
  0 & 0 & \alpha_{121} & 0 & 0 & 0 & 0 & 0 \\
   0 & \alpha_{112} & 0 & 0 & 0 & 0 & 0 & 0 \\
  0 & 0 & 0 & 0 & 0 & 0 & 0 & \alpha_{222}\\
\end{array}%
\right).
\end{equation}
Then we have the following lemma for three-qubit states.
\begin{lem} If elements of $\mathcal{R}$ satisfies
$\mathrm{T}_{i}\neq0$, for $1\leq i\leq 12$, then the state
$\mathcal{R}(\ket{\psi}\otimes\ket{\psi}\otimes\ket{\psi})$, with
$\ket{\psi}=\ket{1}+\ket{2}$ is entangled.
\end{lem}
The proof of this lemma follows by construction of $\mathcal{R}$
which is based on separable elements of three-qubit states defined
by $T_{i}$. For example
$\mathcal{R}(\ket{\psi}\otimes\ket{\psi}\otimes\ket{\psi})=\sum^{2}_{k_{1},k_{2},k_{3}=1}
\alpha_{k_{1}k_{2}k_{3}}\ket{k_{1}k_{2}k_{3}}$ is entangled if and
only if $T_{i}\neq0$.  Note that we can also write the braiding
operator $\mathcal{R}$ for three-qubit as
$\mathcal{R}_{2^{3}\times 2^{3}}=\mathcal{R}^{d}_{2^{3}\times
2^{3}}+\mathcal{R}^{ad}_{2^{3}\times 2^{3}}$, where
$\mathcal{R}^{a}_{2^{3}\times
2^{3}}=(\alpha_{111},0,\ldots,0,\alpha_{222})$ is a diagonal
matrix and $\mathcal{R}^{ad}_{2^{3}\times
2^{3}}=(0,\alpha_{221},\alpha_{212},\ldots,\alpha_{112},0)$ is an
anti-diagonal matrix. We will use this notation to construct the
matrix $\mathcal{R}$ for multi-qubit state.
 For $m$-qubit state a topological unitary
transformation $\mathcal{R}_{2^{m}\times 2^{m}}$ that create
multi-qubit entangled states is defined by
$\mathcal{R}_{2^{m}\times 2^{m}}=\mathcal{R}^{d}_{2^{m}\times
2^{m}}+\mathcal{R}^{ad}_{2^{m}\times 2^{m}}$, where
$\mathcal{R}^{a}_{2^{m}\times
2^{m}}=(\alpha_{1\cdots1},0,\ldots,0,\alpha_{2\cdots2})$ is a
diagonal matrix and $\mathcal{R}^{ad}_{2^{m}\times
2^{m}}=(0,\alpha_{22\cdots1},\ldots,\alpha_{21\cdots1},\alpha_{12\cdots2},
\ldots,\alpha_{1\cdots12},0)$ is an anti-diagonal matrix.
 Then we have following lemma for general multi-qubit
states.
\begin{lem} Let  $\mathcal{X}^{j}_{2\times 2^{m-1}}$ be a $2\times 2^{m-1}$
matrix defined by
\begin{equation}
\mathcal{X}^{j}_{2\times 2^{m-1}}=
\left(%
\begin{array}{cccc}
  \alpha_{11\ldots11_{j}1\ldots1} & \alpha_{11\ldots11_{j}1\ldots2} & \ldots & \alpha_{22\ldots21_{j}2\ldots2}\\
  \alpha_{11\ldots12_{j}1\ldots1} & \alpha_{11\ldots12_{j}1\ldots2} & \ldots & \alpha_{22\ldots22_{j}2\ldots2}\\
\end{array}%
\right).
\end{equation}
Then the state $\mathcal{R}_{2^{m}\times
2^{m}}(\ket{\psi}_{1}\otimes\ket{\psi}_{2}\otimes\cdots\otimes\ket{\psi}_{m})$,
with $\ket{\psi}_{j}=\ket{1}+\ket{2}$ is entangled if $2\times 2$
minors of $\mathcal{X}^{j}_{2\times 2^{m-1}}$ not vanishes, for
all $j=1,2,\ldots,m$.
\end{lem}
The proof  follows by construction of $\mathcal{R}_{2^{m}\times
2^{m}}$ which is based on completely separable elements of
multi-qubit states defined by the Segre ideal
$\mathcal{I}^{m}_{\mathrm{Segre}}=\sum^{m}_{j=1}\mathcal{I}_{\mathcal{Q}_{j}\models\mathcal{Q}_{1}\mathcal{Q}_{2}
\cdots\widehat{\mathcal{Q}}_{j}\cdots\mathcal{Q}_{m}}$. That is
the state $ \mathcal{R}_{2^{m}\times
2^{m}}(\ket{\psi}\otimes\ket{\psi}\otimes\cdots\otimes\ket{\psi})=\sum^{2,2,\ldots,2}_{k_{1},k_{2},\ldots,k_{m}=1}\alpha_{k_{1}k_{2}\ldots
k_{m}} \ket{k_{1}k_{2}\ldots k_{m}}$ is entangled if and only if
all $2\times 2$ minors of $\mathcal{X}^{j}_{2\times
2^{m-1}}\neq0$. Note that this operator is a quantum gate
entangler since $\tau_{2^{m}\times 2^{m}}=\mathcal{R}_{2^{m}\times
2^{m}}\mathcal{P}_{2^{m}\times 2^{m}}$ is a $2^{m}\times 2^{m}$
phase gate and $\mathcal{P}_{2^{m}\times 2^{m}}$ is $2^{m}\times
2^{m}$ swap gate. Thus we have succeeded to construct quantum gate entagler
for a general multi-qubit state based on a similar construct of a braiding  operator that
satisfies condition for separability  that is given by definition of the
Segre ideal. This also shows a good relation between
topology, algebraic geometry and quantum theory with application
in the field of quantum computing.

\begin{flushleft}
\textbf{Acknowledgments:} The  author  acknowledges the
financial support of the Japan Society for the Promotion of Science
(JSPS).
\end{flushleft}


\end{document}